\documentclass{iau}
\usepackage{graphicx}

\title[Galaxy evolution through AGN surveys] 
{AGN surveys to study galaxy evolution along cosmic times}

\author[Luigi Spinoglio]   
{Luigi Spinoglio}

\affiliation{Istituto di Astrofisica e Planetologia Spaziali \\ Via Fosso del Cavaliere 100,
00133, Roma, Italy \\ email: {\tt luigi.spinoglio@iaps.inaf.it}  
}

\pubyear{2013}
\volume{304}  
\pagerange{119--126}
\jname{Multiwavelength AGN surveys and studies}
\editors{A. Mickaelian, F. Aharonian, D. Sanders, eds.}
\begin{document}

\maketitle

\begin{abstract}
Various observational techniques have been used to survey galaxies and AGN, 
from X-rays to radio frequencies, both photometric and spectroscopic. 
I will review these techniques aimed at the study of galaxy evolution and of the role of
AGNs and star formation as the two main energy production mechanisms. 
I will then present as a new observational approach the far-IR spectroscopic surveys 
that could be done with planned astronomical facilities of the next future, 
such as SPICA from the space and CCAT from the ground.
\end{abstract}

\firstsection 
\section{Introduction}

Active Galactic Nuclei (AGN) have been first considered as exotic objects, then it has been 
realised that their occurrence is fundamental in galaxy evolution.

Surveys of AGN have been made in two main different ways, at various wavelengths. The first way has been to
detect their continuum emission at X-rays; in the UV; in the mid-IR and in the radio. The second approach was based
on the detection of spectral emission lines in the UV/optical and in the near- and mid-IR.

For understanding galaxy evolution, one has to study the inter-play of the role of star formation and AGN: 
this can be done only with spectroscopic surveys. 
The best wavelengths are those long enough to minimise the effect of dust extinction, 
which is known to be relevant at least at redshifts of z=1-3.

In a paper of  1946 \cite[(Seyfert 1946)]{seyf46}, entitled "Emission in the nuclei of spiral galaxies", 
appeared three years after his discovery in 1943 \cite[(Seyfert 1943)]{seyf43}, Carl Seyfert wrote that:
"The observed line intensities [....] in NGC1068 and 4151 closely resemble the line intensities of the 
planetary nebula NGC7027" and "The hydrogen lines in NGC4151 and 7469 are of unusual interest,
being composed of relatively narrow cores (1100 km/sec wide) superposed on very wide wings 
(7500 km/sec wide)Ó. So, since their discoveries, Seyfert galaxies were known to have typical signatures
in their optical spectra. After thirty years, in a review paper Weedman concluded with "[.....] However, if
a 10$^7$ black hole or even 10$^7$ neutron stars were inserted into a galactic nucleus, most of the activity
in Seyfert nuclei could be made to happen as a consequence of accretion onto these objects. But an 
explanation of how the required compact objects get into the nuclei requires comprehensive knowledge
of galactic evolution, and that's another story" \cite[(Weedman 1977)]{weed77}.
The necessity to place the Seyfert galaxies phenomenon in the framework of galaxy evolution was 
already clear more than 30 years ago.
\vspace{-5mm}
\section{Continuum and emission line surveys to discover AGN}

I will review some of the most important results, without being exhaustive, of the various continuum and line emission surveys
to discover and identify AGN at various wavelengths.
\vspace{-2mm}
\subsection{Continuum emission}
\begin{itemize}
\item \textbf{UV excess}: 
\cite[Markarian \& Lipovetski (1967-1972)]{mar67, mar67a, mar69, mar69a, mar69b, mar69c, ml71, ml71a, ml72, ml72a}
published data on 500 galaxies selected for having a strong or moderate 
UV continuum on objective-prism plates. Spectroscopic observations of $\sim$400 of these galaxies exhibit emission lines. 
40 objects have well defined spectra typical of Seyfert galaxy nuclei \cite[(Sargent 1970, Khachikian \& Weedman 1974, Arakelian 1975)]{sar70,kw74,a75}.
	
\cite[Green, Schmidt \& Liebert (1986)]{gsl86} published the Palomar Green (PG) catalog of UV-excess stellar objects,  
which is a statistically complete sample of hot stars and quasars, made of digitized and calibrated 
2-color photographic plates (1874 objects from 266 fields, observed with the Palomar 18" Schmidt telescope) 
and (for U-B$<$-0.46) 3700-6500\AA~ spectra at the Hale 5m and 1.5 telescopes.

Malkan \& Sargent (1982) measured the  3000\AA~bump (see Fig 1.a) interpreted as thermal emission at 20-30.000K due to the 
black hole accretion disk.

\begin{figure}[h]
\begin{center} 
 \includegraphics[width=3.2in]{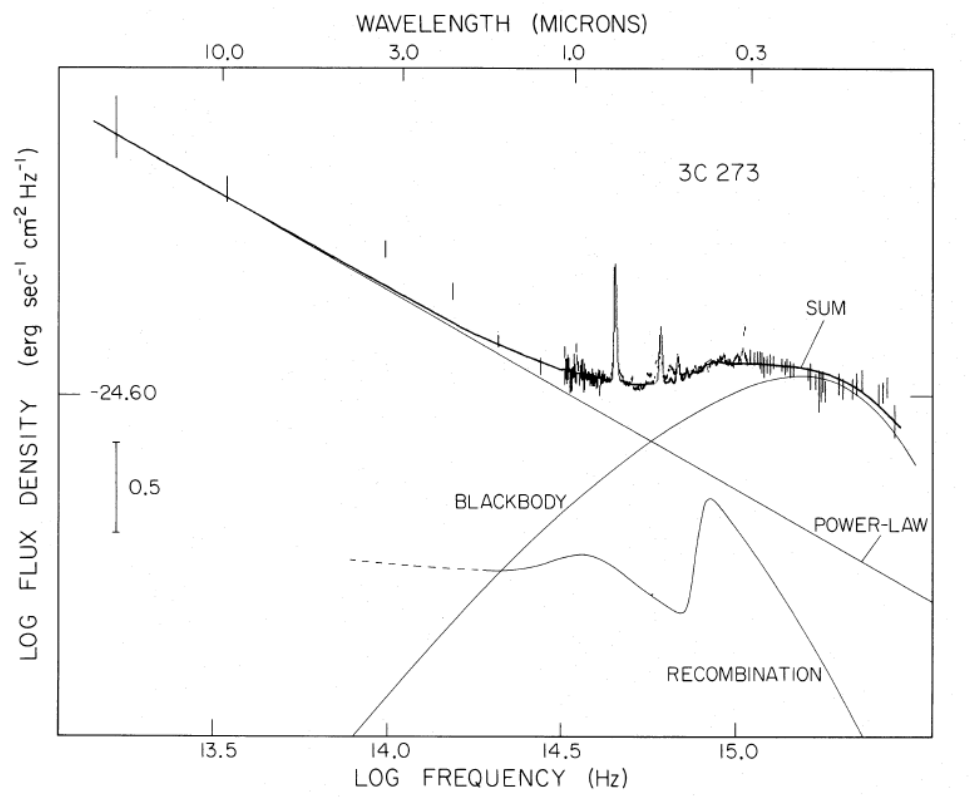}  \includegraphics[width=1.7in]{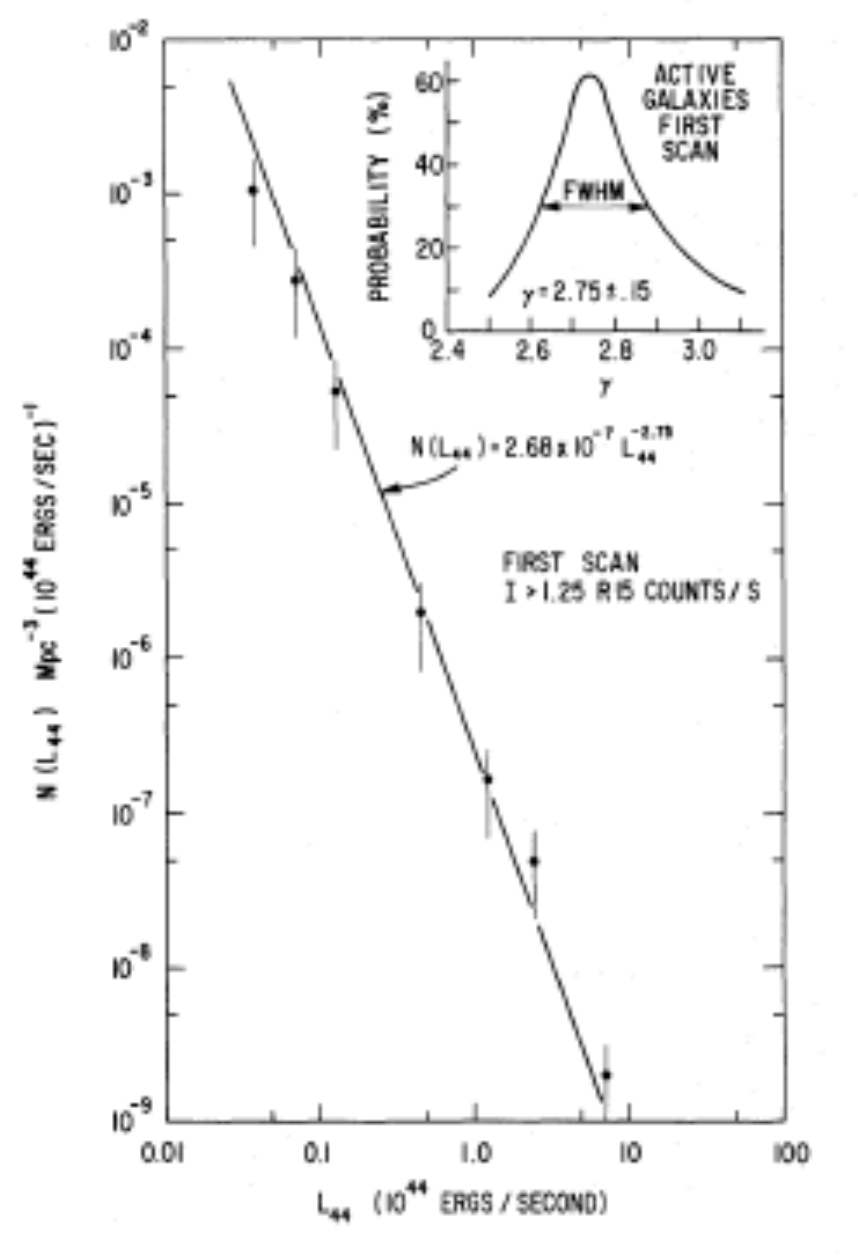} 
 \caption{{\bf Left: a)} Decomposition of the best fit to 3C273: the solid lines show the power law, hydrogen recombinations and black body at 26,000K \cite[(Malkan \& Sargent 1982)]{ms82}. {\bf Right: b)} First determination of the hard X-ray Seyfert galaxy luminosity function \cite[(Piccinotti et al. 1982)]{pic82}. } 
   \label{fig1}
\end{center}
\end{figure}

\item \textbf{Hard X-ray emission}: \cite[Piccinotti et al. (1982)]{pic82}, provided the first complete catalog of 61 X-ray (2-10keV) emitting extragalactic 
objects down to a limit 
of $\sim$ 3.1$\times$10$^{-11}$ ${\rm erg~cm^{-2}~s^{-1}}$. They derived the first X-ray luminosity function
of AGN (see Fig. 1b). 

\cite[Sazonov et al (2007)]{saz07}  have used the INTEGRAL all-sky hard X-ray (17-60keV) survey to perform 
a statistical study of a representative sample of nearby AGN. 
The entire all-sky sample consists of 127 AGN, of which 91 are confidently detected ($>$5$\sigma$) on the time-averaged 
map obtained with the IBIS/ISGRI instrument. 
Among these 
there are 66 non-blazar AGN located at $\mid$b$\mid$$>$5 $\deg$, where the surveyÕs identification 
completeness is $\sim$93\%, which were used for calculating the AGN luminosity function and X-ray absorption distribution.
\cite[Malizia et al. (2012)]{mal12} have studied the fraction of Compton-thick sources in an INTEGRAL complete AGN sample.
Assuming as a dividing line between absorbed and unabsorbed AGN a column density of N$_{H}$=10$^{22}$ cm$^{-2}$, they find that
half (48\%) of the sample is absorbed while the fraction of Compton thick AGN is small (7\%). 
They also showed that these fractions suffer from a bias towards heavily absorbed objects, which are lost if weak and at large distance. 
When this bias is removed, as is possible in the local Universe, the above fractions increase to 80\% and 17\% (within 60 Mpc).

\cite[Tueller et al. (2008)]{tue08} presented the 
analysis of the first 9 months of data of the Swift BAT survey of AGNs in the 14-195 keV band. 
Using archival X-ray data or follow-up Swift XRT observations, 129 (103 AGNs) objects have been 
identified (out of 130) detected at $\mid$b$\mid$$>$15 $\deg$ and with significance $>$4.8$\sigma$.
Integration of the luminosity function gives a local volume density of AGNs above 10$^{41}$ erg/s of 2.4$\times$10$^{-3}$ Mpc$^{-3}$, 
which is about 10\% of the total luminous local galaxy density above M* =-19.75.


\item \textbf{Mid-IR excess}: 
\cite[Spinoglio \& Malkan (1989)]{sm89} first found that there is a spectral interval (7-12$\mu$m) at 
which the absorption of the original continuum is balanced by the thermal emission, because dust absorbs 
the continuum at short wavelengths and re-emit it in the FIR. This balance allows the non-thermal emission of
AGN (power-law) to dominate the continuum in the mid-IR spectral region. As can be seen in 
Fig. 2a, around the wavelength of 12$\mu$m is emitted a constant fraction of bolometric flux for any type of 
active galaxies ($\sim$1/5 of the bolometric flux), therefore the 12$\mu$m selection is equivalent to select at a bolometric flux limit. On Fig. 2b, the first determination of the 12$\mu$m luminosity function of Seyfert galaxies is
shown \cite[(Rush, Malkan \& Spinoglio 1993)]{rms93}.
At higher redshifts, this method has also been applied, e.g. using the {\it Spitzer} observations at 24$\mu$m for
galaxies at redshift of z$\sim$1 and $\sim$2 to detect the rest-frame emission at 12 and 8 $\mu$m, respectively 
\cite [(e.g. Lacy et al 2013)]{lac13}.

\end{itemize}

\begin{figure}[h]
\begin{center} 
 \includegraphics[width=2.65in]{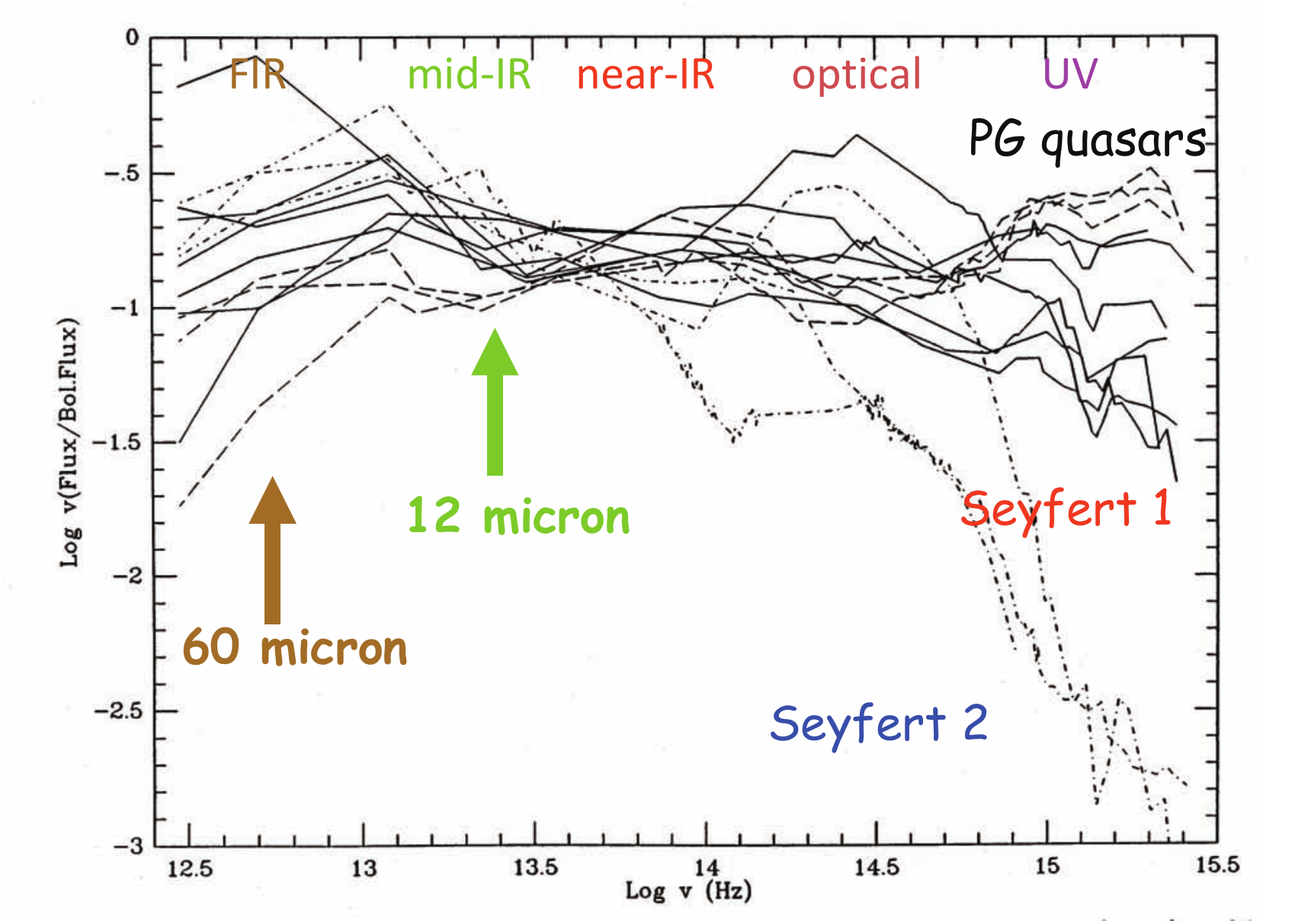}  \includegraphics[width=2.55in]{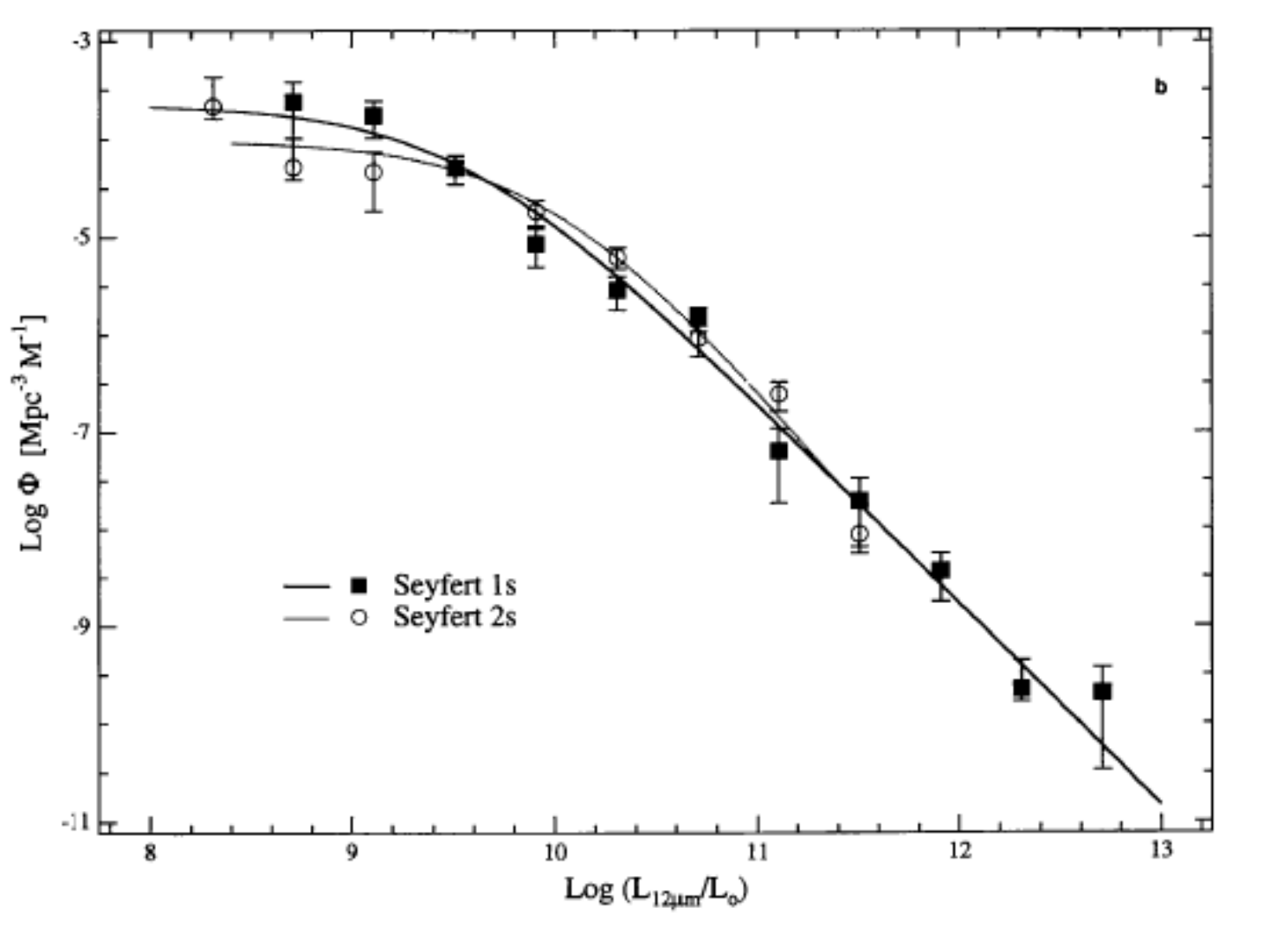} 
 \caption{{\bf Left: a)} Spectral energy distributions of 13 AGN normalized to the bolometric fluxes (computed from 0.1-100$\mu$m) \cite[(Spinoglio \& Malkan, 1989)]{sm89}
. {\bf Right: b)} First determination of the 12$\mu$m Seyfert galaxy luminosity function \cite[(Rush et al. 1993)]{rms93}. } 
   \label{fig1}
\end{center}
\end{figure}

\vspace{-5mm}

\subsection{Emission line spectra}
\begin{itemize}
\item \textbf{Optical spectroscopy}: 
The first complete optically selected sample of  Seyfert galaxies (25 Seyfert 1 and 23 Seyfert 2) has been extracted 
from the CfA Redshift Survey of $\sim$2400 objects (1/50 or 2\% efficiency) \cite[(Huchra \& Burg 1992)]{hb92}.
\cite[Veilleux \& Osterbrock 1987]{vo87}  proposed the popular classification method to classify emission-line galaxies 
through their optical spectra (see Fig. 3a, the classification of CfA active galaxies).
	

The Sloan Survey SDSS \cite[(Eisenstein et~al. 2011, and references therein)]{eis11} 
has covered $\sim$1/3 of the sky in visible-light imaging and spectroscopy and found significant numbers of high-z quasars 
\cite[(Richards etÊal. 2002)]{ric02}. The large amount of data can
be appreciated by looking at the diagnostic diagram of Fig. 3b, where a sample of $\sim$ 23,000 galaxies has been analysed to study metallicities \cite [(Groves et al. 2006)]{gro06}.

	
\end{itemize}

\begin{figure}[h]
\begin{center} 
 \includegraphics[width=2.55in]{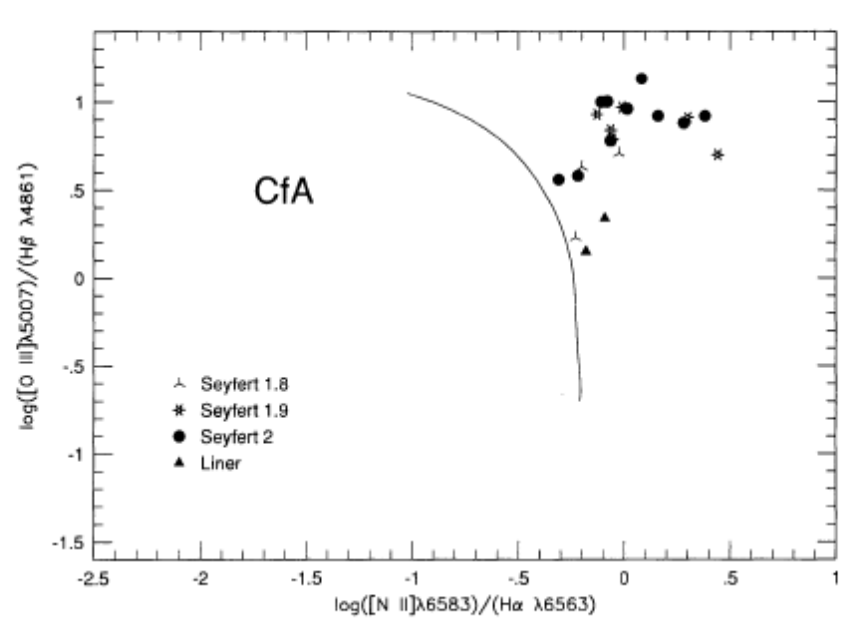}  \includegraphics[width=2.65in]{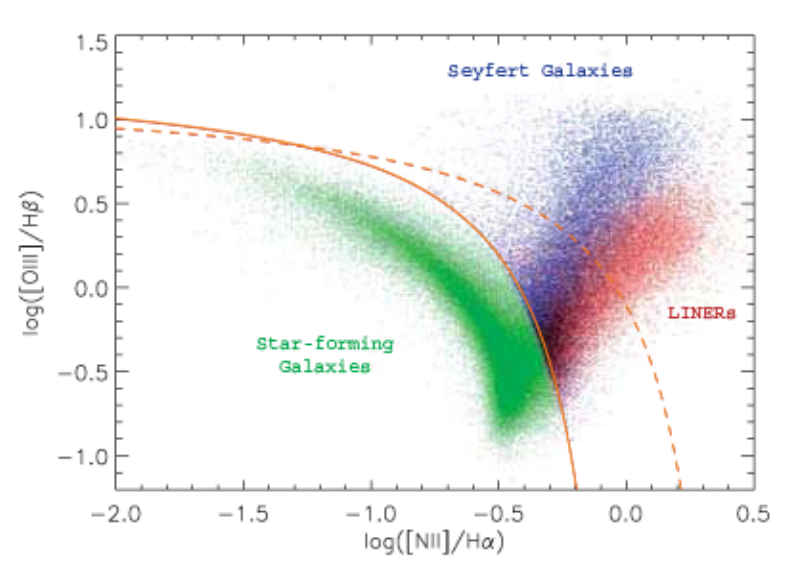} 
 \caption{{\bf Left: a)} reddening corrected [OIII]$\lambda$5007/H$_{\beta}$ vs. [NII]/H$_{\alpha}$ intensity ratios for active
 galaxies in the CfA sample. Solid curve divides AGN from HII region galaxies \cite[(Vielleux \& Osterbrock 1987)]{vo87}. 
 {\bf Right: b)} The same diagram, twenty years later, for active and star-forming galaxies of the Sloan Survey (SDSS) \cite[(Groves et al. 2006)]{gro06}. } 
   \label{fig1}
\end{center}
\end{figure}

\vspace{-3mm}

\section{Galaxy evolution: role of AGNs and Star Formation in galaxies}

To study galaxy evolution we need to study the occurrence and evolution of both AGN and star formation (SF),
because both phenomena deeply influence the history of galaxies. We also know that they are deeply connected,
as shown by the so-called \textit{Magorrian relation} \cite[(Magorrian et al. 1998)]{mag98}. 
We want to know the full cosmic history of energy generation by stars (star formation) and AGNs (black hole accretion). 
These energy production rates correspond to built up the mass (of central black hole, or galactic stars), and must--ultimately--be consistent. We also want to uncover how much of this is partly or heavily extinguished by dust and seek cosmic connections between galaxy's stars and its massive Black Hole: understand the how and why of these systems. 
First, we need to find the right technique to separate the two emission regimes.
\vspace{-2mm}
\subsection{Comparing different techniques for separating AGN and SF}

No single criteria can be used to distinguish AGN and SF, but there are limits and potentialities of different observational techniques:\\
-- UV/Optical/NIR observations are able to measure galaxy morphology and spectra, however they seriously suffer from dust obscuration,
known to be relevant at z$\sim$1--3.\\
-- X-ray observations are good tracers of AGN, however only weak X-ray emission can be detected from star formation and, even more importantly, heavily-obscured AGN (Compton-thick) are completely lost, or partly lost even at the high hard X-rays energies of, e.g., INTEGRAL (see section 2.1).\\
-- Radio observations (with planned facilities like EVLA, SKA) can detect AGN and SF to large z and can see through gas and dust, they can measure morphology and spectral energy distributions (SED), detect polarization and variability, however not always redshifts can be measured. At its highest frequencies, SKA could be able to measure redshifted molecular lines in the ISM of galaxies.\\
-- mm/submm observations (e.g. ALMA, CCAT) will provide spectra from SF (redshifted CO, [CII], etc.), however we need to find AGN tracers at the longest FIR wavelengths. 
-- Rest-frame MIR/FIR imaging spectroscopy can provide a complete view of galaxy evolution by measuring the role of BH and SF because it can (provided that large field of view and high sensitivity can be reached) trace simultaneously both SF and AGN, measure redshifts  and see through large amounts of dust. It seems therefore to be the most promising technique. \\

\vspace{-5mm}

\subsection{The power of infrared spectroscopy}

Fig.4a shows how well the IR fine structure lines cover the density-ionization parameter space which characterizes the photoionized and photon dissociated gas \cite[(see, e.g., Spinoglio \& Malkan 1992)]{sm92}. 
A combination of these lines and line ratios can trace both star formation and black hole accretion. The long wavelengths of these lines, ranging from the far-IR for the photodissociation and HII region lines, through the mid-IR for the AGN lines, to the near-IR for the coronal lines, ensure that we can observe these different tracers by minimising the effect of dust extinction.

The rich rest-frame mid-IR spectra, that have been recently observed in active and starburst galaxies in the local Universe with the mid-IR spectrometer 
onboard the {\it Spitzer} satellite 
can be observed in the far-IR in the redshift range of 0.4$<$z$<$3.0. 

Fig.4b shows the average Spitzer IRS 
mid-IR spectra \cite[(Tommasin et al. 2010)]{tommasin10} of subclasses of Seyfert galaxies from the the 12$\mu$m Seyfert galaxy sample of \cite[Rush, Malkan \& Spinoglio (1993)]{rms93}.  
For comparison, we also show the average spectrum of starburst galaxies \cite[(Bernard-Salas et al. 2009)]{b-s09}. The quality of the data is very high and shows the many features that can distinguish between AGN and star formation processes, such as the high-ionization lines from [NeV] 
originated exclusively from AGN or the 11.2$\mu$m PAH feature and the low ionization lines from [NeII] and [SIII], typical of HII and star forming regions. Mid-/far-IR imaging spectroscopy is therefore able to trace galaxy evolution throughout cosmic times in an unbiased way by minimising dust extinction.

\begin{figure}[h]
\begin{center} 
 \includegraphics[width=2.6in]{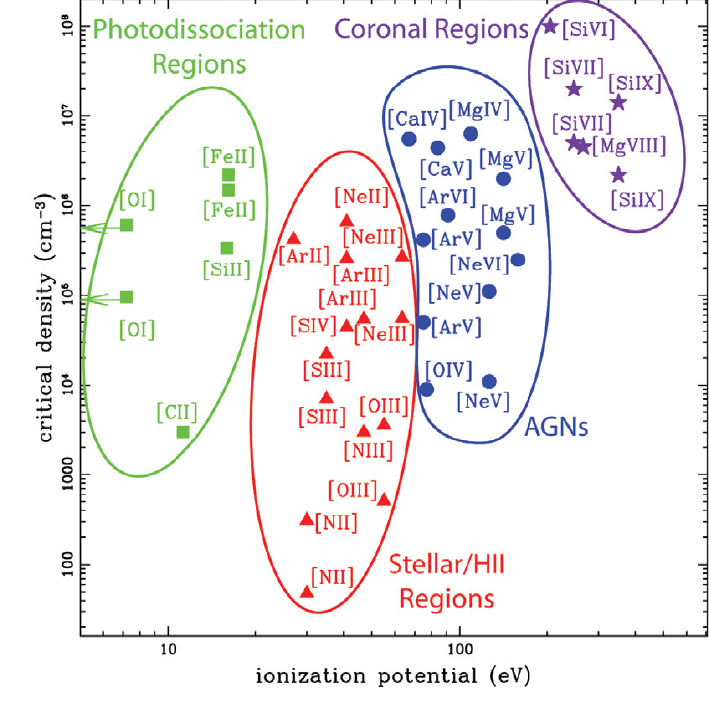}  \includegraphics[width=2.6in]{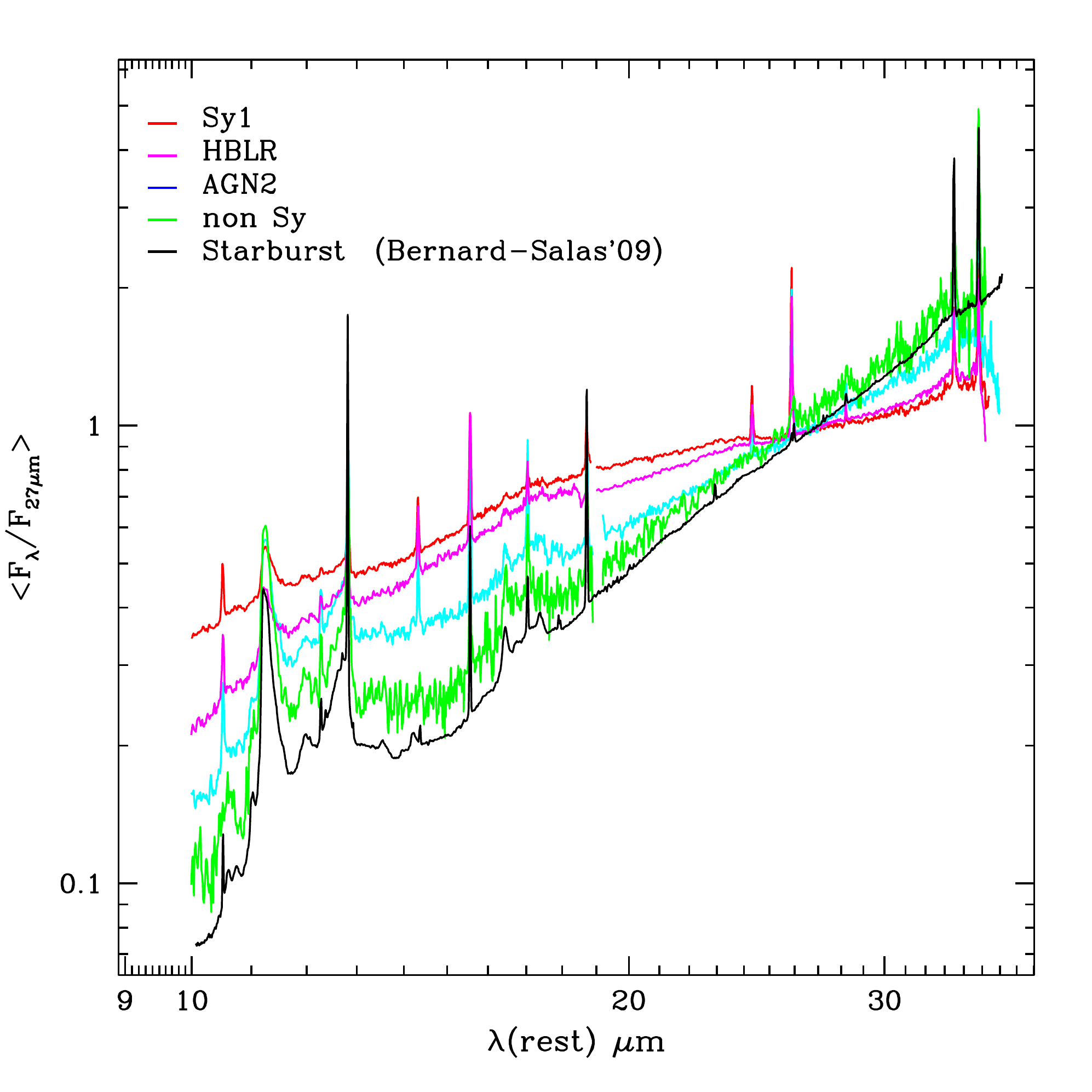} 
 \caption{
 {\bf Left: a)}  Critical density for collisional de-excitation vs. ionization potential of IR fine-structure lines, showing the diagnostic 
 power of these lines to trace different astrophysical conditions 
 \cite[(Spinoglio \& Malkan 1992)]{sm92}.
 {\bf Right: b)}  Mid-IR spectra of Seyfert galaxies 
 normalized at 27$\mu$m, showing a sequence with decreasing 
 non-thermal activity, from Seyfert  1's through Hidden Broad Line Region (HBLR) galaxies and type 2 AGN to low luminosity 
 AGN (non-Seyfert's) \cite[(Tommasin et al. 2010)]{tommasin10}, compared to 
 starburst galaxies \cite[(Bernard-Salas et al. 2009)]{b-s09}. 
}
   \label{fig2}
\end{center}
\end{figure}

\vspace{-3mm}


Due both to the atmospheric absorption, which leaves open only a few sparse windows in the near- and mid-IR, and to
the high thermal background at room temperature at IR wavelengths, it has soon been realised that infrared astronomy 
to be successful had to be done from space telescopes, as it was demonstrated by the success of the various space missions, from
IRAS  \cite[(Neugebauer et al. 1984)]{Neugebauer_1984} to {\it Herschel} \cite[(Pilbratt et al. 2010)]{Pilbratt_2010}.  
However, due to the sensitivity limits and the poor multiplexing power of the spectrographs onboard of these spacecrafts, 
only a few limited samples of distant objects have been successfully observed \cite[(e.g., Yan et al. 2007; Men{\'e}ndez-Delmestre et al.
2009; Sturm et al. 2010)] {yan07,men09,stu10}, 
while most of the spectroscopic work has been done in the local Universe.
Substantial progress in studying galaxy evolution therefore can only be achieved by using direct mid- to far-IR spectroscopic surveys, which will provide measured (rather than estimated) redshifts and also unambiguously characterise the detected sources, by measuring the AGN and starburst contributions to their bolometric luminosities over a wide range of cosmological epochs, through their spectroscopic signatures.

SPICA \cite[(Nakagawa et al. 2011)]{Nakagawa_2011} will be the next-generation, space infrared observatory, which, for the first time, 
will contain a large (3.2-meter) actively cooled telescope (down to 6K), providing an extremely low background 
environment. With its instrument suite, 
SPICA will provide not only high spatial resolution and unprecedented sensitivity in mid- and far-infrared imaging, but especially
large field medium spectral resolution imaging spectroscopy. These characteristics put SPICA among the best planned facilities to
perform spectroscopic cosmological surveys in the mid- to far-IR. Using theoretical models for galaxy formation and evolution constrained by the luminosity 
functions observed with both {\it Spitzer} and {\it Herschel} and the relations between line and continuum far-IR
luminosity, as measured in the local Universe for active and starburst galaxies, \cite[Spinoglio et al. (2012)]{spinoglio12} have predicted, as a function of redshift, the intensities of key lines able to trace AGN
and star formation activity along cosmic history.

\vspace{-3mm}

\begin{figure}[!ht]
\begin{center}
   \resizebox{1.\hsize}{!}{
     \includegraphics*{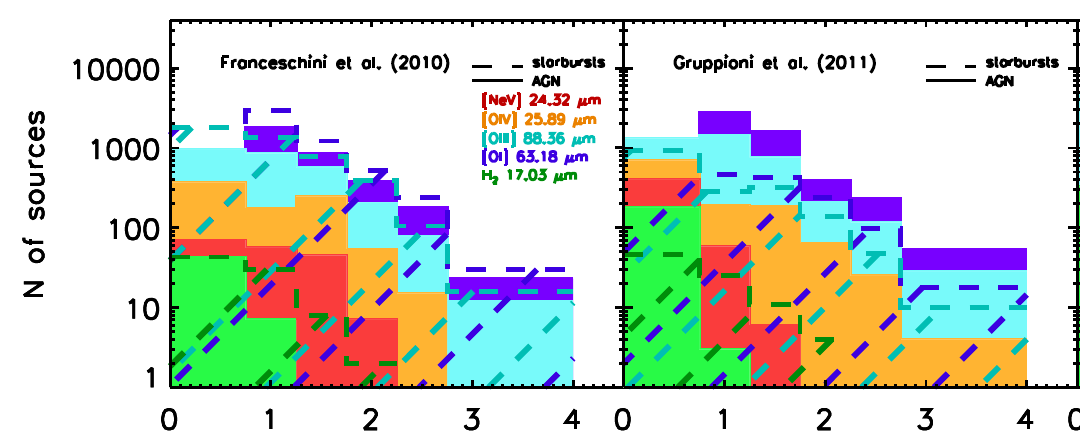}
   }
\end{center}
\caption{Predicted number of objects as a function of redshift, per spectral line and object type, (AGN are shown as continuum lines, starburst
 galaxies as dashed lines) in 1 hr. integration per pointing, 0.5 $deg^{2}$ survey with the SAFARI instrument \cite[(Roelfsema et al. 2012)]{roel12} requiring 450 hrs. of total integration time, following two different models \cite[(Spinoglio et al. 2012)]{spinoglio12}. 
}
\label{fig:fig3}
\end{figure}


Figure~\ref{fig:fig3} shows graphically the number of galaxies that can be detected by the far-IR spectrometer SAFARI \cite[(Roelfsema et al. 2012)]{roel12} planned to be onboard of SPICA, in each spectral line for the two different populations of AGN 
and starburst 
galaxies, comparing 
two models. The total numbers of detectable objects agree, taking the different models, to within a factor of 2-3. 
At least a thousand galaxies will be simultaneously detected in four lines at 5$\sigma$ over a half square degree.  A survey of the given assumptions will lead to the detection of bright lines (e.g., [O I] and [O III]) and PAH features in thousands of galaxies at z$>$1. Hundreds of z$>$1 AGN will be detected in the [O IV] line, and several tens of z$>$1 sources will be detected in [Ne V] and H$_2$.

On the other hand, the Cerro Chajnantor Atacama Telescope (CCAT) \cite[(Sebring et al. 2010)]{seb10} will be 
complementary to SPICA, being able to observe the [OIII]88$\mu$m line at z $>$ 1.3.
We also find that CCAT will be very efficient 
for studies of [CII], an important coolant of the interstellar medium (ISM), at all z $<$ 5. At 3 $<$ z $<$ 4 alone, it will detect more than 300 galaxies at 5$\sigma$ level in a 0.5 deg$^2$ survey \cite[(Spinoglio et al. 2012)]{spinoglio12}.

\section{Conclusions}
We summarise this work with these points:
\begin{itemize}
\item[-]  After many decades of efforts, we are close to having reliable measures of SF rate and AGN accretion power, through MIR/FIR spectroscopic surveys, unaffected by dust. 
\item[-]  Accurately measuring the star formation rate and the AGN accretion power is the first step towards understanding galaxy evolution over the history of the Universe.
\item[-] FIR spectroscopic surveys with SPICA will physically measure galaxy evolution.
\item[-] Given the expected sensitivity of  SAFARI $\sim$2.5$\times$10$^{-19}$ W/m$^2$ (5$\sigma$,1 hr.) thousands of sources will be detected in more than 4 lines in typical 0.5 deg$^2$ surveys (t=450 hrs.).
\item[-] Complementary to SAFARI, CCAT will detect several hundreds of galaxies at R$\sim$1000 in a 0.5 deg$^2$ survey in 4.5 hours in  [OIII]88$\mu$m and thousands of galaxies in [CII]158$\mu$m.
\item[-]These surveys will be essential to clarify the inter-relation between quasar activity and star formation, which of the two processes influence the other and ultimately will test the processes able to shape the mass and luminosity functions of galaxies.
\end{itemize}

\end{document}